# A Partially Supervised Bayesian Image Classification Model with Applications in Diagnosis of Sentinel Lymph Node Metastases in Breast Cancer

Ying Zhu[1,*], Tom Fearn[2], D.Wayne Chicken[3], Martin R. Austwick[3], Santosh K. Somasundaram[3], Charles A. Mosse[3], Benjamin Clark[3], Irving J. Bigio[4], Mohammed R.S. Keshtgar[5], Stephen G. Bown[3]

[1]*Mathematics and Mathematics Education, Nanyang Technological University, Singapore*

[2]*Department of Statistical Science, University College London, UK*

[3]*Research Department of Tissue & Energy, Division of Surgery & Interventional Science, University College London, UK*

[4]*Department of Biomedical Engineering and Department of Electrical & Computer Engineering, Boston University, Massachusetts, USA*

[5] *Division of Surgery and Interventional Sciences, University College London, UK*

September 2017

# ABSTRACT

A method has been developed for the analysis of images of sentinel lymph nodes generated by a spectral scanning device. The aim is to classify the nodes, excised during surgery for breast cancer, as normal or metastatic. The data from one node constitute spectra at 86 wavelengths for each pixel of a 20×20 grid. For the analysis, the spectra are reduced to scores on two factors, one derived externally from a linear discriminant analysis using spectra taken manually from known normal and metastatic tissue, and one derived from the node under investigation to capture variability orthogonal to the external factor. Then a three-group mixture model (normal, metastatic, non-nodal background) using multivariate $t$ distributions is fitted to the scores, with external data being used to specify informative prior distributions for the parameters of the three distributions. A Markov random field prior imposes smoothness on the image generated by the model. Finally, the node is classified as metastatic if any one pixel in this smoothed image is classified as metastatic. The model parameters were tuned on a training set of nodes, and then the tuned model was tested on a separate validation set of nodes, achieving satisfactory sensitivity and specificity. The aim in developing the analysis was to allow flexibility in the way each node is modelled whilst still using external information. The Bayesian framework employed is ideal for this.

**Key words:** Image classification; Discriminant dimension reduction; Principal component analysis; Model-based clustering; Bayesian multivariate finite mixture model; Markov random field.

## 1. Introduction and background

When breast cancer spreads to other parts of the body it does so via the chain of axillary lymph nodes in the armpit. If the first node in this chain, the sentinel node, is clear of metastases, then the remaining nodes are almost certainly clear also (Keshtgar and Ell, 2002). Thus, a rapid diagnostic method that enables an excised sentinel node to be checked during surgery can be used to avoid the unnecessary removal of all the nodes or to avoid the need for repeat surgery at a later date, following positive determination of metastases. The currently available methods, touch imprint cytology and frozen section histopathology, require the presence of an expert pathologist, something that is not always feasible. The molecular diagnostic technique like One Step Nucleic Acid (OSNA) test (Huxley *et al.*, 2015) although reliable, is time consuming and requires expensive equipment with high running cost. An alternative optical diagnostic method, elastic scattering spectroscopy (ESS), has been proposed, and has shown considerable promise (Keshtgar *et al.*, 2010; Austwick *et al.*, 2010).

In the device that motivated the research described in this paper, elastic scattering spectra in the wavelength range 320-800 nm are automatically measured on a $20 \times 20$ grid over the cut surface of an excised sentinel node. After smoothing and thinning, the spectrum for each grid element had $p = 86$ data points (wavelengths) at just under 5-nm intervals. The challenge is to develop an algorithm that will use the grid of 400 spectra from one node to classify the node as metastatic or normal, according to whether it contains any metastatic tissue or not.

A method for achieving this has been reported by Keshtgar *et al*. (2010) and Austwick *et al*. (2010). The first step was to derive a linear discriminant function (LDF) from an analysis of around 3000 individual spectra measured on completely normal and completely metastatic nodes. These measurements used the same fibre-optic probe and spectrometer as in the new device, but pre-dated the construction of the automated system and were taken manually at selected points on the nodes. To classify a scanned node, any pixels lying outside the node, typically in the corners of the grid, are removed by visual inspection, and then each of the remaining pixels is classified as normal or metastatic using the LDF. Finally, to avoid an unacceptable number of false positives arising from the misclassification of individual pixels, a node is declared to be metastatic only if there is a cluster of at least 9 contiguous pixels, all classified as metastatic.

This method was shown to work well (Austwick *et al*., 2010), but there may be scope for improving on it. One drawback was the need for manual intervention to remove non-nodal areas from the image, another was that the final image of classified pixels was not convincing, because of misclassified pixels. The cluster-of-9 rule solves this problem regarding classification, but does not clean up the image. Finally, there is considerable variability among nodes, and possibly some slight mismatch between the training data from manually measured spectra and the spectra from the new device.

The approach described here attempts to tackle these problems. The basic method is model-based clustering (Fraley and Raftery, 2002), applied to each node. The manual training data are used to define one of the dimensions in the low-dimensional space, onto which the spectra are projected for this clustering, and to determine prior distributions for the parameters of the normal and metastatic groups. The idea is that the manual training data should strongly guide the clustering, whilst still leaving some scope for different solutions for different nodes, hence the description "partially supervised". There is an additional cluster for the non-nodal region, allowing this to be detected automatically. After the clustering has produced an initial solution, that solution is used as the starting point for fitting a hidden Markov random field model (Geman and Geman, 1984; Li, 2001), which exploits the spatial structure of the data to produce a smoother and much more plausible image, as well as doing away with the need for the cluster-of-9 rule.

This report is organized as follows: Section 2 introduces the instrumentation system and describes the ESS data; Section 3 describes the proposed two-stage partially supervised image classification model; Section 4 describes the application to the diagnosis of sentinel lymph nodes metastases, and there are some further discussions and conclusion in Section 5.

## 2. Instrumentation system and data description

The study of the feasibility of using ESS to discriminate between normal lymph nodes and those containing metastatic tissue was conducted in two phases. In the first phase, a total of $m = 3{,}213$ spectra from 339 normal nodes and 30 totally metastatic nodes were collected using a hand-held probe. Each spectrum was measured by placing the optical probe manually at up to 16 sites on the cut surface the bisected node. Each node, and therefore each site, was classified as normal or metastatic by histopathology experts.

For the second phase, an automated two-dimensional ESS scanning device was constructed. The ESS scanner instrumentation, shown in Figure 1, consists of a pulsed xenon arc lamp, a static ESS fibre-optic probe, a mobile sample stage, a spectrometer, and a computer to control the various components and to record the spectra. The cut surface of the node is placed under a square fibre-optic plate, and the device moves the node and plate under the probe to take measurements at all points on a $20 \times 20$ grid. This covers the cut surface of the node and in most cases also includes some non-nodal areas, possibly contaminated by blood or lipid and varying considerably from node to node. No training set from the manual data was available for this non-nodal group. The result is a $20 \times 20$ image with $n = 400$ pixels, each of size $0.5 \times 0.5$ mm, and a full ESS spectrum for each pixel. For each node a photograph with a microscopic view is taken of the area being scanned. These photos will be used later to compare with the images generated by our model. We have used data from 117 nodes including 65 normal and 52 with metastases. As before, each node was classified as normal or metastatic by histology experts, but this classification is not available for individual pixels in the image. Apart from the problem with non-nodal pixels, most of these metastatic nodes include pixels of normal nodal tissue.

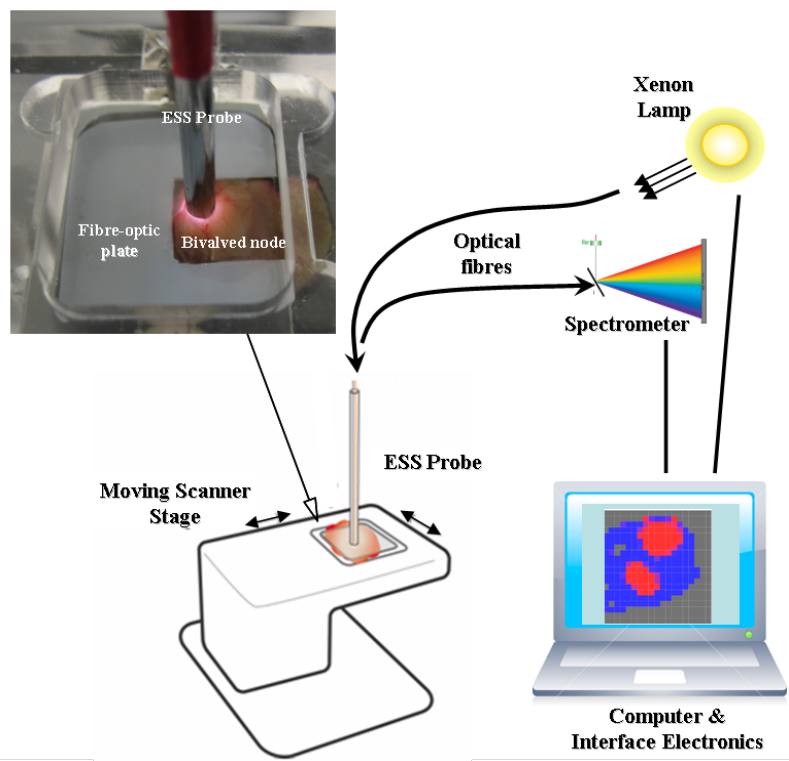

Figure 1: Schematic diagram of elastic scattering spectroscopy (ESS) scanning device system.

## 3. Two-stage partially supervised image classification model

### 3.1 Discriminant dimension reduction and variable construction

Some dimension reduction of the spectral data is essential to enable the feasibility of the multivariate distribution fitting. Given that the eventual aim is real-time prediction, reduction to a small number of dimensions is desirable.

The proposed approach, called discriminant dimension reduction, takes as the first dimension, which we call the external variable, the canonical variate derived from a linear discriminant analysis (LDA) on the manually measured data. This axis gives maximum separation between normal and metastatic groups in the manual data. To allow the method to adapt to each individual node, a small number of internal variables derived from a principal component analysis (PCA) of the variability orthogonal to the external variable in the spectra of the node under study, are added to this external variable.

*Step 1: Constructing the external variable*

A principal component discriminant analysis (PCDA), PCA followed by a LDA, was carried out on the manual measurement data, $X_{train}$, with the first $k_{ext}$ principal components being used in the LDA to construct a scalar canonical variable that separates the normal and metastatic groups. The choice of $k_{ext}$ was made by leave-out-one-site cross-validation on the manual data. The scores on this variable are calculated as

$$T_{train} = X_{train}\, q_{ext}, \tag{1}$$

where each element of the $m \times 1$ vector $T_{train}$ is the canonical score for one spectrum in the manual training data, $X_{train}$ is an $m \times p$ matrix of spectra, and $q_{ext}$ is a $p \times 1$ loading vector from the PCDA. In computing these scores, $X_{train}$ was centred. The same centring, i.e., using the mean of $X_{train}$, was applied to other spectral matrices when computing the scores in expressions (2), (4) and (6) later.

By applying the external variable loading, $q_{ext}$, derived from the manual data, to the spectral data from the node of interest, we compute the external variable for this node

$$T_{node.ext} = X_{node}\, q_{ext} \tag{2}$$

where each point of the $n \times 1$ vector $T_{node.ext}$ is a scalar canonical score $t_{node.ext}$, and $X_{node}$ is an $n \times p$ spectral matrix for the scanned node of interest, with each row of $X_{node}$ being a $p$-dimensional spectrum for one pixel in the image.

*Step 2: Constructing the internal variable(s)*

To construct the internal variables, the first $k_{int}$ PCs of the node of interest are extracted from the space orthogonal to the external variable, thus

$$\tilde{X}_{node} = X_{node}(I - q_{ext}(q_{ext}^T q_{ext})^{-1} q_{ext}^T) \quad (3)$$

$$T_{node.int} = \tilde{X}_{node} Q_{int} \quad (4)$$

where $I$ is an $p \times p$ identity matrix, $\tilde{X}_{node}$ is an $n \times p$ spectral matrix whose columns lie in the $p-1$ dimensional subspace orthogonal to the external variable, $Q_{int}$ is a matrix, the columns of which are the first $k_{int}$ principal component loadings of $\tilde{X}_{node}$, and $T_{node.int}$ is an $n \times k_{int}$ score matrix. Each row of $T_{node.int}$ is $k_{int} \times 1$ internal variable, $t_{node.int}$, which is composed of the first $k_{int}$ PCs in the subspace orthogonal to $q_{ext}$. The data vector used in the later data analysis is $x = (t_{node.ext}, t_{node.int})$, with a dimension of $k = 1 + k_{int}$.

The manual measurement data, $X_{train}$, are then projected onto the same subspace as for the node of interest, using the loadings from expressions (3) and (4) by using

$$\tilde{X}_{train} = X_{train}(I - q_{ext}(q_{ext}^T q_{ext})^{-1} q_{ext}^T) \quad (5)$$

and

$$T_{train.int} = \tilde{X}_{train} Q_{int}, \quad (6)$$

which converts $X_{train}$ into an $m \times k_{int}$ score matrix $T_{train.int}$. The $k \times 1$ means, $m_n, m_c$, and $k \times k$ covariance matrices, $V_n$, $V_c$, of the $k$ variables $(T_{train}, T_{train.int})$ are calculated from the two groups of normal and metastatic data respectively. These will be used in the priors in the mixture model described below.

By using the external variable we impose one dimension from the training data which we believe can separate normal from metastatic tissue. Adding internal variables retains the variability specific to this node.

## 3.2 Stage 1: Bayesian model-based clustering

### 3.2.1 Multivariate *t* mixture model

#### 3.2.1.1 The model

Suppose $x_1,..., x_n$ are $k$-dimensional random observations generated independently from a mixture of $g$ underlying populations (groups) so that

$$f(x_i | \phi) = \sum_{j=1}^{g} \pi_{ij} f(x_i | \theta_j) \qquad (7)$$

where $f(x_i | \theta_j)$ denotes the conditional probability density function of $x_i$, belonging to the *j*th group, parameterized by $\theta_j$, and each $\pi_{ij}$ is the probability that pixel *i* belongs to group *j*, which we allow to depend on the position in the image as described below. Here $\phi = (\pi_{i1},..., \pi_{ig}, \theta_1,..., \theta_g)$ denotes the set of unknown parameters. For this application, we assume that $g = 3$ with normal, metastatic and non-nodal groups, and $x_i$ is the dimension-reduced $k$-variate spectral data, measured at pixel *i* of the image for one node. Here we use the multivariate *t* distribution, that is, $x_i | \theta_j \sim t(\mu_j, \Sigma_j, v_j)$, because this provides a more robust approach to the fitting of mixture models than the use of normal components and gives less extreme estimates of the posterior probabilities of component membership of the mixture models, as demonstrated in Peel and McLachlan (2000). In general it is possible to estimate the $v_j$, but we will fix them, and in fact we use the same $v_j$ for different groups, so that the parameters to be estimated are $\theta_j = \{\mu_j, \Sigma_j\}$. In the two stages, different values for $v_j$ are allowed, with $v_{s1}$ and $v_{s2}$ used for stage 1 and stage 2, respectively.

#### 3.2.1.2 Priors for parameters $\theta_j$

For the finite mixture model a normal inverse Wishart prior (Gelman *et al.,*1995) is used here as a prior for $\theta_j$. Instead of using a scalar as a common prior weight for all the dimensions in the normal inverse Wishart prior (Fraley and Raftery, 2007), an extended normal inverse Wishart prior is developed here by defining a diagonal matrix allowing different prior weights in different dimensions.

The prior on the mean vector $\mu_j$, conditional on the scale matrix $\Sigma_j$ is taken as

$$\mu_j | \Sigma_j \sim N(\mu_{jp}, K_{jp}^{-1/2} \Sigma_j K_{jp}^{-1/2}), \qquad (8)$$

where $\mu_{jp}$ is a $k$-dimensional vector, and the $K_{jp}$ is a $k \times k$ diagonal matrix with diagonal elements (scalars) $\kappa_{jp_1},\ldots,\kappa_{jp_k}$ of prior weights.

The scale matrices $\Sigma_j$ are given inverse Wishart priors

$$\Sigma_j \sim IW(v_{jp}, \Lambda_{jp}), \tag{9}$$

where $v_{jp}$ is a scalar, and $\Lambda_{jp}$ is a matrix of the same dimension as $\Sigma_j$, as suggested in Raftery (1996).

The joint prior $p(\theta_j)$ is therefore

$$p(\mu_j, \Sigma_j \mid \mu_{jp}, K_{jp}, v_{jp}, \Lambda_{jp}) \sim N(\mu_{jp}, K_{jp}^{-1/2} \Sigma_j K_{jp}^{-1/2}) \cdot IW(v_{jp}, \Lambda_{jp}) \tag{10}$$

The hyperparameters $\mu_{jp}, K_{jp}, v_{jp}$ and $\Lambda_{jp}^{-1}$ ($j$ = 1, 2, 3 for normal, metastatic and non-nodal components, respectively), are called the mean, prior weight, degrees of freedom and scale, respectively, of the prior distribution. Here the suffix $p$ is used to indicate a hyperparameter and is not the number of wavelength points of the spectrum. The choices for the hyperparameters will be discussed in Section 4.1.

### 3.2.1.3 Priors for parameters $\pi_{ij}$

Because the background component is more likely to appear on the fringe area of the node, we use a position parameter $\alpha_{ij}$ defined by a background score $\omega_i$, to allow the probability of the pixel being a background component to depend on its position in the image. We begin by defining a scaled Euclidean distance of a pixel from the centre of the image,

$$d_i = d(r_i, s_i) = \frac{\sqrt{(r_i - r_c)^2 + (s_i - s_c)^2}}{9.5\sqrt{2}}, \; r_i \in \{1,2,\ldots,20\}, \; s_i \in \{1,2,\ldots,20\} \tag{11}$$

where $(r_i, s_i)$ denotes the row position (index) and the column position (index) of the pixel $i$ in the image, and $(r_c, s_c) = (10.5, 10.5)$ is the centre of the image. The scaled distance $d_i$ is dimensionless and it varies from 0.0526 for the most central pixels to 1 for those in the corners.

As a function of $d_i$, $\omega_i$ gives a background score to each pixel as follows:

$$\omega_i = f(d_i) = \begin{cases} \min((d_i)^{1/\rho}, 0.97), & \text{if } d_i > 0.56 \\ d_i & \text{otherwise} \end{cases} \tag{12}$$

where the power $1/\rho$ is used to emphasize the scores for the pixels on the edge, and the 0.97 prevents the probability from being too close to 1. The threshold of 0.56 (the scaled radius of a circle that reaches to two pixels from the edge of the image) is used to define the fringe area of the image.

Using the background score, $\omega_i$, the position parameter $\alpha_{ij}$ is defined to allow the probability of the pixel being background ($j = 3$) to depend on its position in the image, with the remaining probability being split equally between the two nodal groups.

$$\alpha_{ij} = \begin{cases} \omega_i, & \text{if } j = 3 \\ (1-\omega_i)/2, & \text{if } j = 1, 2 \end{cases} \quad (13)$$

where, as stated earlier, the coding is 1, 2 and 3 for normal, metastatic and non-nodal components, respectively.

Combining this with a factor $\pi_j$, representing the abundance of component $j$, the probability of a pixel $i$ being the $j$th component is denoted by

$$\pi_{ij} = \alpha_{ij}\delta_i\pi_j, \quad (14)$$

where $\delta_i = (\sum_j \alpha_{ij}\pi_j)^{-1}$ is the normalization factor that leads to $\sum_j \pi_{ij} = 1$.

Using this scheme, the probability of pixels on the corner or the edge of the scanned area being a background component is much higher than the probability for those in the centre.

### 3.2.2 EM for mixtures of multivariate $t$ distributions via Bayesian theory

To fit the $g$-component mixture of multivariate $t$ distributions to the scanned data, where class membership is unknown for individual pixels, we introduce the membership indicator variable, $z_i = (z_{i1},...,z_{ig})$, whose role is to encode the component that has generated the $i$th observation. This will be treated as missing data. The indicators $z_i$ ($i = 1,...,n$) are a set of binary variables $z_{ij} \in \{0,1\}$ ($j = 1,...,g$) with

$$z_{ij} = \begin{cases} 1 & \text{if observation } i \text{ belongs to group } j \\ 0 & \text{otherwise} \end{cases}, \quad (15)$$

and, hence, $\sum_{j=1}^{g} z_{ij} = 1$.

If we know $z_i$, we can write

$$f(x_i | z_i, \phi) = \prod_{j=1}^{g} \{f(x_i | \theta_j)\}^{z_{ij}} . \tag{16}$$

The indicator variable $z_i$ is multinomial with probabilities $\pi_{ij}$, so that the joint density of $x_i$ and $z_i$ is given by

$$f(x_i, z_i | \phi) = \prod_{j=1}^{g} \{\pi_{ij} f(x_i | \theta_j)\}^{z_{ij}} . \tag{17}$$

Here, $f(x_i | \theta_j)$ is a multivariate $t$ distribution $t(\mu_j, \Sigma_j, v_j)$. To facilitate the computations with this distribution, a set of weights, $\{u_i, i=1,...,n\}$, are introduced, one corresponding to each of the observations $x_i$, so that

$$x_i | u_i, z_{ij} = 1 \sim N(\mu_j, \Sigma_j / u_i), \tag{18}$$

and

$$u_i | z_{ij} = 1 \sim G(v_j/2, v_j/2) \tag{19}$$

independently for $i=1,...,n$. Integrating $u_i$ would give the original distribution $t(\mu_j, \Sigma_j, v_j)$. The $u_i$'s are also treated as missing data.

We use an expectation-maximization (EM) algorithm to estimate the unknown parameters $\phi = \{\theta, \pi\}$ with $\pi = \{\pi_j\}$ (McLachlan and Krishnan, 1997). The complete data are $(\{x_i\}, \{z_i\}, \{u_i\})$. At this stage the class membership of pixel $i$ in the image is assumed independent of all the other pixels, without considering the spatial correlation of the image. Then the complete-data log-posterior distribution is

$$\ell(\phi | \{x_i\}, \{z_i\}, \{u_i\}) = \sum_{i=1}^{n} \sum_{j=1}^{g} z_{ij} \{\log f(x_i | u_i, \theta_j) + \log f(u_i | v_j) + \log \pi_{ij}\} + \sum_{j=1}^{g} \log p(\theta_j) , \tag{20}$$

where $p(\theta_j)$ is the extended normal inverse Wishart prior distribution $p(\mu_j, \Sigma_j | \mu_{jp}, K_{jp}, v_{jp}, \Lambda_{jp})$ for the parameters $\theta_j$ in Equation (10), and the $\pi_{ij}$ is as defined in Equation (14), with a uniform prior for the $\pi_j$. The parameters $\hat{\theta}_j^{(0)} = \{\hat{\mu}_j^{(0)}, \hat{\Sigma}_j^{(0)}\}$ are initialized by using a single-link hierarchical clustering analysis based on Euclidean distances between objects. The vector $\hat{\pi}^{(0)}$ is initialized by giving equal proportions in the mixture. In general it is possible to estimate $v_j$, but in our application to mixture models, we will take $v_j$ as fixed.

The EM algorithm iterates from these starting values, alternating between E and M steps. The E step evaluates the conditional expectation of the complete-data log-posterior density over $\{z_i\}$ and $\{u_i\}$ given the observed data $\{x_i\}$, and the current parameter estimates $\hat{\phi}^{(t)}$ (Zhu, 2009). In this case the result is to replace $z_{ij}$ and $u_{ij}$ in (20) by estimates.

At the ($t$+1)st iteration, the updating equation for $\hat{z}_{ij}$ is given by

$$\hat{z}_{ij}^{(t+1)} = \frac{\hat{\pi}_{ij}^{(t)} f(x_i | \hat{\theta}_j^{(t)})}{\sum_{j=1}^{g} \hat{\pi}_{ij}^{(t)} f(x_i | \hat{\theta}_j^{(t)})} \quad , \tag{21}$$

where $\hat{z}_{ij}$ is the conditional probability that pixel $i$ belongs to the $j$th component, given data $x$ and the current parameter estimates $\hat{\phi}^{(t)}$.

The updating equation for $\hat{u}_{ij}^{(t+1)}$ is given by

$$\hat{u}_{ij}^{(t+1)} = \frac{v_j + k}{v_j + (x_i - \hat{\mu}_j^{(t)})^T \hat{\Sigma}_j^{(t)-1} (x_i - \hat{\mu}_j^{(t)})} . \tag{22}$$

The M step involves maximizing the log posterior distribution over $\pi_j$ and $\theta_j$ with $z_{ij}$ and $u_{ij}$ substituted by their current estimates. At the ($t$+1)st iteration, let $\hat{n}_j^{(t+1)} = \sum_{i=1}^{n} \hat{z}_{ij}^{(t+1)}$, then we have

$$\hat{\pi}_j^{(t+1)} = \frac{\hat{n}_j^{(t+1)}}{\sum_{i=1}^{n} \hat{\delta}_i^{(t+1)} \alpha_{ij}} \quad , \tag{23}$$

with $\hat{\delta}_i^{(t+1)} = (\sum_j \alpha_{ij} \hat{\pi}_j^{(t+1)})^{-1}$, so that $\sum_j \alpha_{ij} \hat{\delta}_i^{(t+1)} \hat{\pi}_j^{(t+1)} = 1 \; \forall i$. An iteration between $\delta_i$ and $\pi_j$ is needed here.

Given $\Sigma_j$, let $\hat{n}_{uj}^{(t+1)} = \sum_{i=1}^{n} \hat{z}_{ij}^{(t+1)} \hat{u}_{ij}^{(t+1)}$, $\hat{\bar{x}}_j^{(t+1)} = \frac{1}{\hat{n}_{uj}^{(t+1)}} \sum_{i=1}^{n} \hat{z}_{ij}^{(t+1)} \hat{u}_{ij}^{(t+1)} x_i$, and $w_{0j} = K_{jp}^{1/2} (\hat{\Sigma}_j^{(t+1)})^{-1} K_{jp}^{1/2}$, then the estimate for $\mu_j$ is:

$$\hat{\mu}_j^{(t+1)} = [w_{0j} + (\hat{\Sigma}_j^{(t+1)})^{-1} \hat{n}_{uj}^{(t+1)}]^{-1} [w_{0j} \mu_{jp} + (\hat{\Sigma}_j^{(t+1)})^{-1} \hat{n}_{uj}^{(t+1)} \hat{\bar{x}}_j^{(t+1)}] \tag{24}$$

Given $\mu_j$, and letting $S_j^{(t+1)} = \sum_{i=1}^{n} \hat{z}_{ij}^{(t+1)} \hat{u}_{ij}^{(t+1)} (x_i - \hat{\mu}_j^{(t+1)})(x_i - \hat{\mu}_j^{(t+1)})^T$, the estimate for $\Sigma_j$ is:

$$\hat{\Sigma}_j^{(t+1)} = \frac{\Lambda_{jp}^{-1} + K_{jp}^{1/2} (\hat{\mu}_j^{(t+1)} - \mu_{jp})(\hat{\mu}_j^{(t+1)} - \mu_{jp})^T K_{jp}^{1/2} + S_j^{(t+1)}}{v_{jp} + \hat{n}_j^{(t+1)} + k + 2} \tag{25}$$

Since (24) involves $\hat{\Sigma}_j$ and (25) involves $\hat{\mu}_j$, a few iterations between these equations are necessary.

The whole EM algorithm is judged to have converged when the relative changes in all elements of $\theta_j$ and $\pi_j$ are less than $\varepsilon$. Here we choose the value 0.01 for $\varepsilon$, in order to achieve a weak convergence in this stage, which merely provides a starting configuration for the second stage. The class label $y_i$ is estimated by $\hat{y}_i = \arg\max_j \hat{z}_{ij}$, when the 1$^{st}$ stage convergence is reached.

### 3.3 Stage 2: Partially supervised Bayesian imaging classification with Markov random field prior

The model in this stage takes into account the spatial correlation in the image by adding a Markov random field spatial prior to the previous model. Since the property of neighbourhood contiguity of the image is now considered, the classification model framework aims at generating an image with smooth pattern, comparable to the real tissue structure of the node.

### 3.3.1 Markov random field spatial prior

We assume that the true configuration $y$ is a realization of a locally dependent Markov random field (MRF). Following the suggestions of Besag (1986), we model the conditional prior probability of pixel $i$ having class label $j$, given the class labels of all other pixels, in the following way:

$$\pi_{ij} = P(y_i = j \mid y_{\partial i}) \propto \alpha_{ij} \exp\{-\beta \, \gamma_{ij}(y)\} \qquad (26)$$

where $\partial i$ is the set of neighbours of $i$, and $\gamma_{ij}(y)$ is the proportion of neighbours having class memberships different to $j$. In our model we always consider a second-order neighbourhood, that is, the eight pixels surrounding each single pixel of the image.

In Equation (26), $\beta$, taken as fixed, is a smoothness parameter capturing the strength of the association between neighbouring pixels, which, when positive, discourages neighbours having different labels. The position parameter $\alpha_{ij}$ is defined by Equation (13) as in stage 1.

## 3.3.2 Parameter estimation in MRF model

A restoration maximization (RM) algorithm (Qian and Titterington, 1991) is used here for fitting the multivariate $t$ mixture model with MRF prior.

The joint probability of the spectral observations $x$ and pixel labels $y$ is

$$f(x, y \mid \theta) = f(x \mid y, \theta) p(y \mid \alpha, \beta) \quad (27)$$

where $p(y \mid \alpha, \beta)$ is the joint density of the MRF model, corresponding to a Gibbs distribution generated by the conditional probabilities defined in Equation (26), $\alpha = \{\alpha_{ij}\}$ are the position parameters for the pixels in the image, and $\theta = \{\theta_1, ..., \theta_g\}$ are the parameters of the component distributions in the mixture.

Introducing the binary membership indicator variable, $z_i = (z_{i1}, ..., z_{ig})$, and the weight variable $u_i$ in multivariate $t$ distribution, as in Section 3.2.2, the complete-data log-posterior density function becomes

$$\ell(\theta \mid \{x_i\}, \{z_i\}, \{u_i\}) =$$

$$\sum_{i=1}^{n} \sum_{j=1}^{g} z_{ij} (\log f(x_i \mid u_i, \theta_j) + \log f(u_i \mid v_j)) + \sum_{j=1}^{g} \log p(\theta_j) + \log p(y \mid \alpha, \beta) \quad (28)$$

which reduces to the expression in Equation (20) when the $y_i$'s are independent of each other.

The EM algorithm applied in Section 3.2.2 will not work here. The M-step is straightforward, since we are only estimating $\theta$, and not $\alpha$ or $\beta$, but the E-step requires the $z_{ij}$ in Equation (28) to be replaced by their conditional expectation, given the current parameter estimates and the observed data $x$. This is non-trivial, because of the dependence structure introduced by the MRF.

In order to deal with this, the restoration-maximization (RM) algorithm is used here. This generates a sequence of pairs $\{\theta^{(t+1)}, y^{(t+1)}\}$ such that $y^{(t+1)}$ is updated on the basis of $x$ and $\theta^{(t)}$, and $\theta^{(t+1)}$ is updated from $x$ and $y^{(t+1)}$.

Using the parameter estimate $\hat{\theta}$ and the image configuration $\hat{y}$ from the first stage algorithm as the starting values $\{\theta^{(0)}, y^{(0)}\}$, the general procedure of the RM algorithm is as follows.

(1) **The R-step (the E-like step)** updates $y^{(t+1)}$ from $p(y|x,\theta^{(t)})$.

In order to update each pixel $y_i$, we adopt the approach suggested by Besag (1986) in his iterated conditional modes (ICM) algorithm. Given the data $x$ and the current realization of the neighbourhood $y_{\partial i}$, the algorithm updates each pixel by the class label $y_i$ which maximizes the conditional posterior probability:

$$\hat{z}_{ij}^{(t+1)} = \frac{\hat{\pi}_{ij}^{(t)} f(x_i|\hat{\theta}_j^{(t)})}{\sum_{j=1}^{g} \hat{\pi}_{ij}^{(t)} f(x_i|\hat{\theta}_j^{(t)})} \quad \text{for } j = 1, 2, \ldots, g \text{ and } i = 1, 2, \ldots, n, \qquad (29)$$

where $\pi_{ij}^{(t)} = p(y_i = j | y_{\partial i}) \propto \alpha_{ij} \exp\{-\beta\gamma_{ij}(\hat{y}^{(t)})\}$. \hfill (30)

Then the class labels are updated by

$$\hat{y}_i^{(t+1)} = \arg\max_j \hat{z}_{ij}^{(t+1)}. \qquad (31)$$

(2) The updating equation for $\hat{u}_{ij}^{(t+1)}$ is the same as in Equation 22. **The M-step** updates $\theta^{(t+1)}$ by maximizing with respect to $\theta$ Equation (28) with $z_{ij}$ replaced by $\hat{z}_{ij}$.

We have the same updating equations as in Section 3.2.2 (Equations 24-25) for $\hat{\mu}_j$ and $\hat{\Sigma}_j$.

## 4. Application and results: diagnosis of sentinel lymph node metastases

### 4.1 Model implementation

The image classification model described in Section 3 was applied to the diagnosis of sentinel lymph node metastases in breast cancer from the ESS images. As described in Section 2, two data sets collected in different ways (that is, the ESS manual measurement data and the ESS scanned data), were used in this study. The manual data were used to define one discriminating dimension and to provide prior distributions for the analysis of the nodes. The scanned data from the 117 nodes (including 65 normal and 52 metastases) were randomly split into two sets with half of the nodes in each class being placed in the first set and the remaining nodes in the second set. The first set including 59 nodes, was used as a training set to tune the parameters $\rho$, $K_{jp}$, $k_{int}$, $v_{s1}$, $v_{s2}$ and $\beta$. Then, with these parameters fixed, the second set including 58 nodes was used as an independent test set to test the algorithm.

Before implementing the proposed model, standard data pre-processing was carried out on spectra from both manual measurements and scanning measurements to improve signal quality (Næs *et al.*, 2002; Johnson *et al.*, 2004; Zhu *et al.*, 2009; Austwick *et al.*, 2010). This involved spectral smoothing, using the Savitzky-Golay filter (Savitzky and Golay, 1964), cropping the noisy ends of the spectra below 400 nm and above 800 nm, and normalizing by using the standard normal variate (SNV) method (Barnes *et al.*, 1989).

**4.1.1 Analysis of manual data**

The mean ESS reflectance spectra from the manual measurements, after standard pre-processing, are shown in Figure 2. A preliminary analysis on the manual measurement data was first carried out by a PCA followed by a LDA to find the canonical variate, the direction maximizing the discrimination between normal and metastatic spectra. Leave-out-one-site cross-validation was used to choose the number of principal components $k_{ext}$ to assess the accuracy of the LDA analysis on a per-site basis. Here 20 principal components were chosen for $k_{ext}$ to construct the external variable.

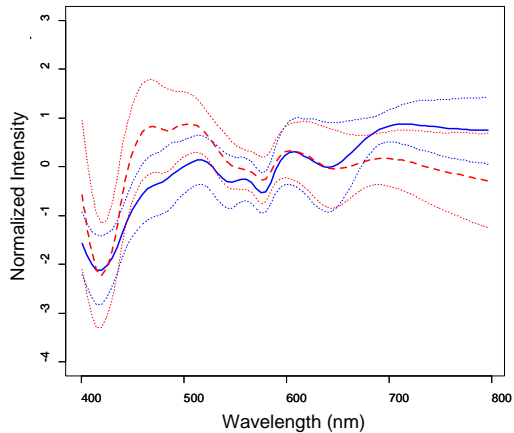

Figure 2: Mean spectra from normal (blue solid line) and metastatic (red dashed line) node with one standard deviation on either side of the mean (dotted lines) after standard pre-processing.

Figure 3 shows the distribution of canonical scores from normal (blue) and metastatic spectra (red) nodes in the manual measurement data derived from the LDA. There are two dominant peaks, the first at a score near 0, corresponding to normal nodes and some fraction of metastatic nodes, and the second at 4 corresponding to metastatic nodes. Scores from metastatic nodes have a broad distribution suggesting that the metastatic areas are genuinely broadly variable (Austwick *et al.*, 2010).

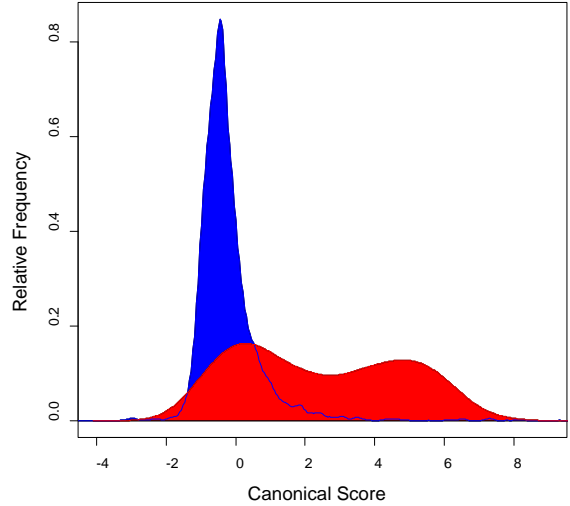

Figure 3: Distribution plot of LDA canonical scores of spectra from manual measurement data. The frequency is plotted as a proportion of class with normal nodes shown in blue and metastatic nodes in red.

The discriminant dimension reduction method in Section 3.1 was used to project the pre-processed spectral data from the scanned lymph nodes into a low-dimensional space composed of one direction of the canonical variate derived from the LDA on the first $k_{ext}$ ($k_{ext} = 20$) PC scores of the manual data, and a few directions derived from the first $k_{int}$ PCs of the variations orthogonal to it, in the spectra of the nodes. The dimension-reduced scanned data of each node thus contain one external variable and one or more internal variable(s), with dimensions reduced from $n \times p$ ($n = 400$, $p = 86$) to $n \times (1 + k_{int})$.

The manual data were then projected onto the space spanned by the directions of the external and internal variable(s) for each node. Along these directions, means and variances derived from normal and metastatic spectra were calculated to be used in the priors for normal and metastatic components.

**4.1.2 Analysis of scanned data**

External and internal variables were constructed on each scanned node for a value of $k_{ext}$, the number of principal components constructing the external variable, fixed at 20, and for a value of $k_{int}$, the number of internal variable(s), ranging from 1 to 5.

A partially supervised image classification algorithm employing a Bayesian multivariate finite mixture model was then applied to the low-dimensional scanned data to model the three unknown groups (normal, metastatic and non-nodal) in the images. Multivariate *t* distributions with degrees of freedom, $v$, ranging from 3 to 20 were tried for the component density of the mixture model. The extended normal inverse Wishart prior with parameters derived from the manual measurements was used as a prior on the parameters of the components of the mixture, with Gaussian prior for component mean vector $\mu$, conditional on scale matrix $\Sigma$, and inverse Wishart prior for $\Sigma$ as in Equation (10). The following choices were made following the suggestions from Fraley and Raftery (2007) for the prior hyperparameters for multivariate mixtures. Here the prior hyperparameters, mean $\mu_{jp}$, scale $\Lambda_{jp}^{-1}$ and prior weight $K_{jp}$ can take different values for each component of the mixture ($j = 1, 2, 3$), and we write $\mu_p = (\mu_{1p}, \mu_{2p}, \mu_{3p})$, $\Lambda_p^{-1} = (\Lambda_{1p}^{-1}, \Lambda_{2p}^{-1}, \Lambda_{3p}^{-1})$ and $K_p = (K_{1p}, K_{2p}, K_{3p})$.

- $\mu_p$ and $\Lambda_p^{-1}$: For the normal and metastatic components, we take $m_n$ and $m_c$, the mean of normal and metastatic groups from the manual data as prior means $\mu_{1p}, \mu_{2p}$, and take $\Lambda_{1p}^{-1} = (v_p - k - 1) V_n$, $\Lambda_{2p}^{-1} = (v_p - k - 1) V_c$ as prior scales. For the non-nodal component, 542 spectra visually recognized by an experienced physicist from some non-nodal areas of the scanned nodes were used to generate a prior mean and a prior scale.

- $v_p$: The marginal prior distribution of $\mu$ is a multivariate *t* distribution centred at $\mu_p$, with $v_p - k + 1$ degrees of freedom. Here we choose $v_p = k + 2$, the smallest integer value for the degrees of freedom that gives a finite covariance matrix (Schafer, 1997), using the same degrees of freedom for all components.

- $K_p$: The posterior mean of group *j* in Equation (24) can be considered as adding $K_{jp}$ observations with value $\mu_{jp}$ to group *j*. For nodal components, a strong prior weight is given on the first dimension of the priors and a weak prior weight is given on the second dimension. For the non-nodal component, a strong prior weight is given on the second dimension as explained in Section 4.3.3. For final model with $k = 2$, the values of diag [5, 2], diag [3, 1.25] and diag [3.85, 10] are taken as $K_{1p}, K_{2p}$ and $K_{3p}$ for the normal, metastatic and non-nodal components, respectively. These specific values were tuned and determined by experiments, which gave the plausible pictures and converged faster. More details of this are given in Section 4.3.3.

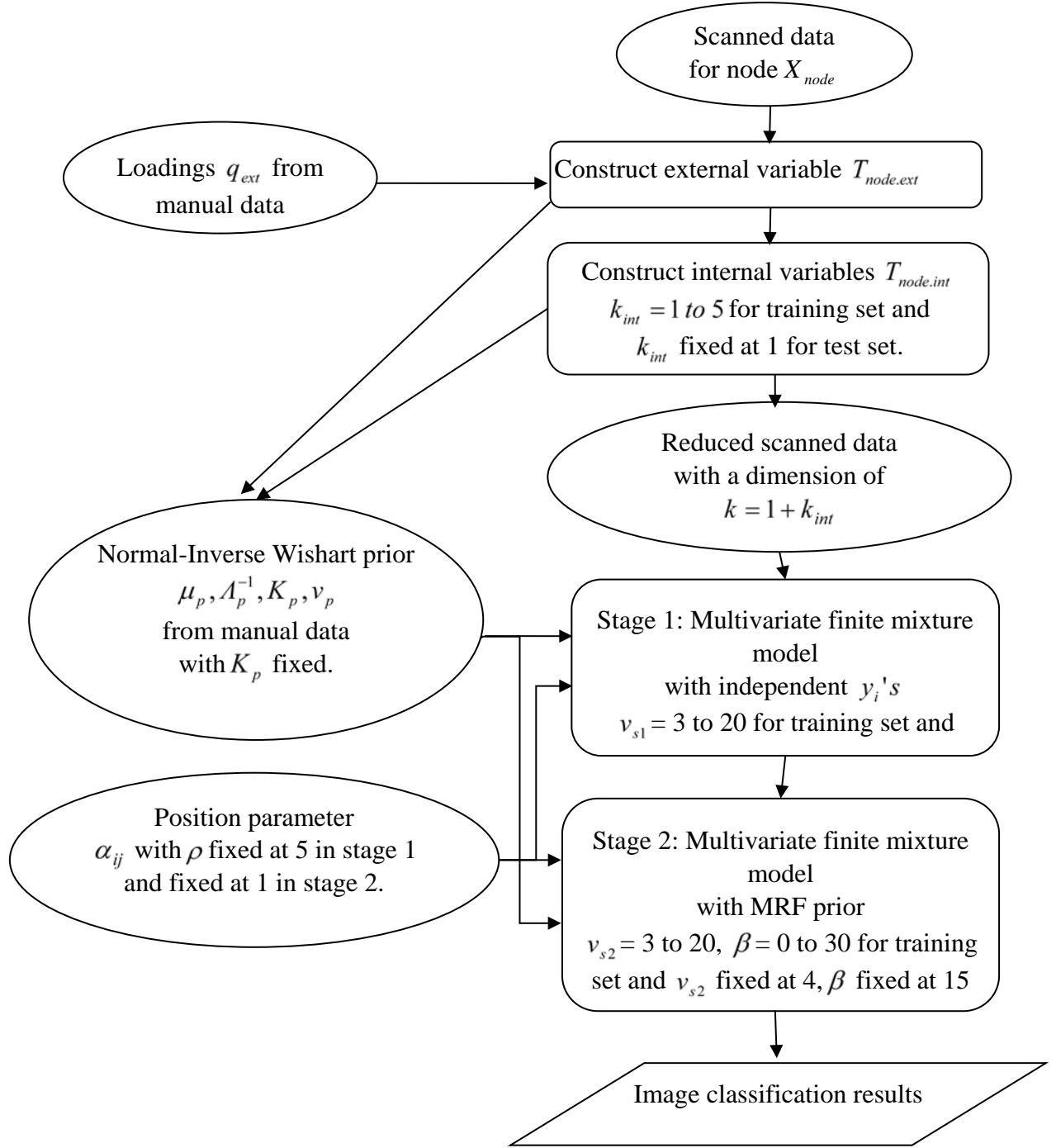

Figure 4: Hierarchical structure of observations, parameters, priors, and values of constants used in our analysis of the sentinel lymph node data. In the rectangular boxes are tuning parameters or choices, in the ellipses are data or fixed constants.

The position parameter $\alpha_{ij}$ as specified in Section 3.2.1.3 varies with the power $\rho$ and $\rho$ is fixed at 5 in stage 1 which is decided by experiments.

The model fitting in stage 1 was implemented by the iterated EM algorithm in Section 3.2.2.

In stage 2 spatial interactions between neighbouring or nearby pixels modelled by a Markov random field prior were then imposed onto the Bayesian multivariate finite mixture model, with smoothing parameter $\beta$ ranging from 0 to 30. The other prior specifications and the form of the position parameter (with $\rho$ fixed at 1 which is a value decided by experiments) were unchanged from the model without spatial interaction.

The model fitting in stage 2 was implemented by the RM (the EM-like) algorithm in Section 3.3.2 starting from the configuration reached after the stage 1 fitting.

A hierarchical structure of the model framework is shown in Figure 4.

## 4.2 Image classification performance assessment

Since reference pathology is not available for individual pixels of the images, but only available for each node, the classification was carried out on a per-node basis. To define conditions for labelling each node as metastatic or non-metastatic, we simply counted the number of positive (metastatic) pixels in the node. The likelihood of scattered false positive pixels occurring over a node is low, since the spatial correlation between adjacent pixels of an image has been taken into consideration in the model fitting. Hence, we classify a node as metastatic if it has even one pixel thus classified.

An image was generated by plotting a $20 \times 20$ matrix of probabilities with the following colour codings. Black indicates the pixels with non-nodal component having the highest probability; for normal or metastatic component we used the posterior probability of the pixel belonging to the metastatic component to generate a colour between red (represents large) and blue (represents small) for each pixel. This image was compared by eye with the photograph of the node to assess the method's success in reconstructing its shape.

In order to search for optimal combinations of $k_{ext}$, $k_{int}$, $v_{s1}$, $v_{s2}$ and $\beta$ (with $\rho$

fixed at 5 in stage 1, and fixed at 1 in stage 2), $k_{ext}$ and $k_{int}$ were first decided by experiment using restricted choices of $v_{s1}$, $v_{s2}$ and $\beta$, and then we fixed $k_{ext} = 20$ and $k_{int} = 1$, the values which gave the best results in this restricted search using the training set of scanned data.

During the experiments for this two-stage image classification model, the combinations of $k_{ext} = 20$, $k_{int} = 1$, $v_{s1} = 4$, $v_{s2} = 4$, and $\beta = 15$ gave the best results, with sensitivity, specificity and AUC of 85%, 94% and 0.91, respectively. The optimal model was applied to the independent test set and gave prediction results with sensitivity of 85% and specificity of 91%. The scanned node spectral data are therefore reduced to a space with two dimensions, one external variable, and one internal variable. Here we focus on the behaviour of this optimal model on an individual node basis, by exploring the dimension reduction method, two-stage model fitting, and model sensitivity to the choices of parameters and priors.

**4.3 Exploring the behaviour of the image classification model**

**4.3.1 How the discriminant dimension reduction method works with the image classification model**

Figure 5 demonstrates how the three types of pixel in a partially metastatic node are classified in a two-dimensional space constructed by discriminant dimension reduction method. The classification of the pixels in the image results from the application of the optimal model.

Using two dimensions in directions orthogonal to each other seems to be a reasonable and sufficient choice for discrimination among the three groups. The two dimensions (i.e. the external and internal variables) typically work as two classifiers. The external variable (in the left panel of Figure 5) discriminates between the two nodal (metastatic and normal) components, and the internal variable (in the right panel of Figure 5) captures the remaining features of an individual node. The mapping image derived from the data in a reduced two-dimensional space is close to the real picture of the node.

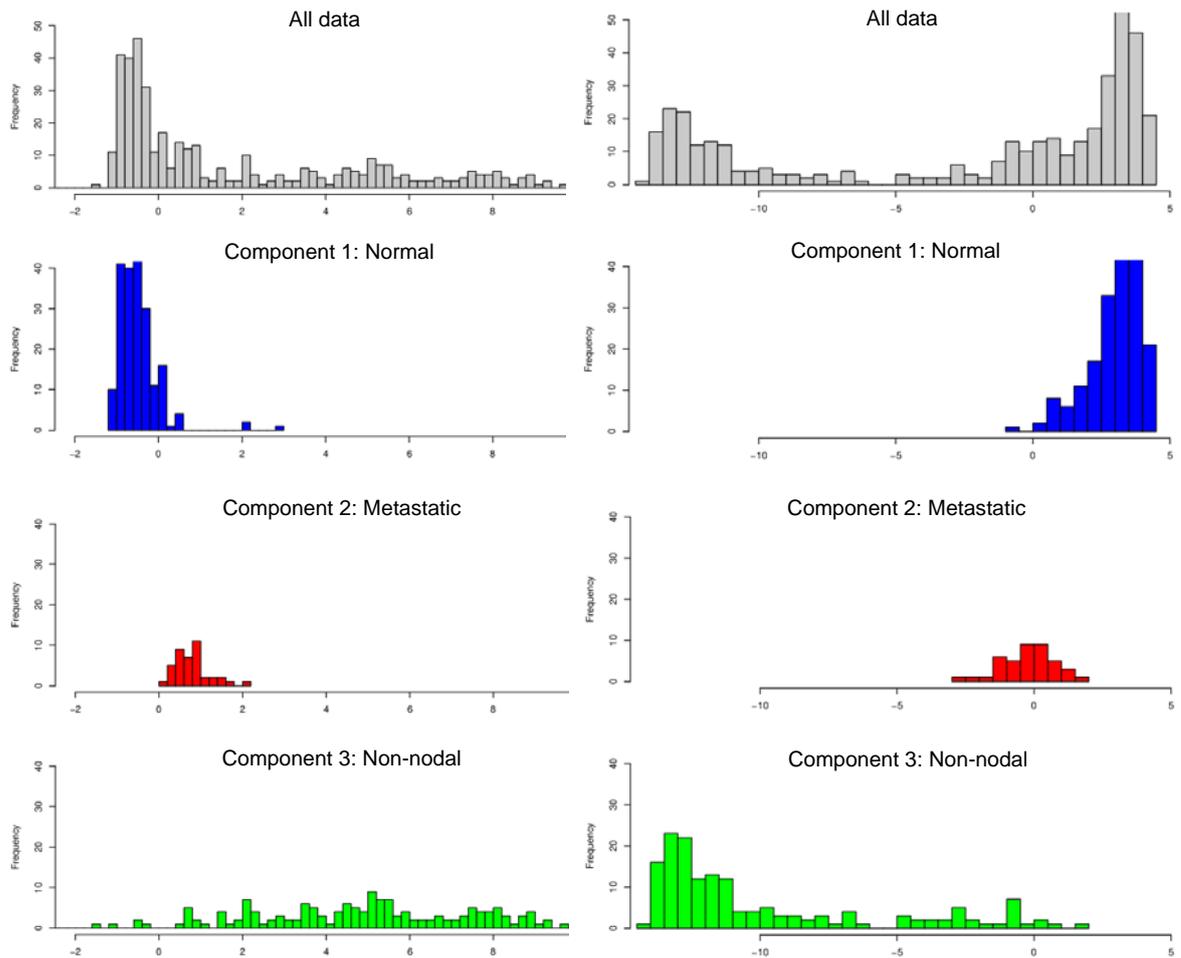
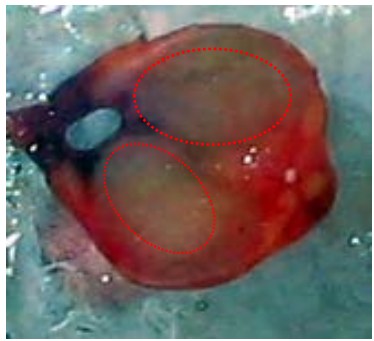
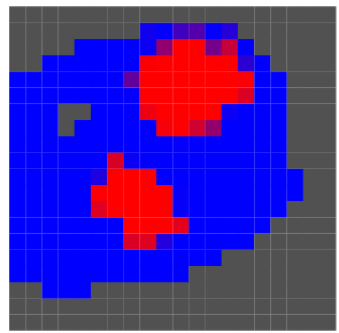

Figure 5: Plots of external variable scores (left panel) and internal variable scores (right panel) of spectra from a partially metastatic node after dimension reduction. In rows 2, 3, 4 histogram of normal spectra (i.e. spectra classified as normal) is shown in blue, metastatic in red and non-nodal in green. The last row shows the photograph (left) and the constructed image (right, red indicative of metastatic spectra, blue of normal and black of non-nodal) of this node.

### 4.3.2 How the two-stage model works for image classification

Figure 6 shows the two-stage image classification for dimension-reduced data of a partially metastatic node. In the left panels, points in blue, red and green refer to the pixels classified as normal, metastatic and non-nodal component, respectively, at the current stage.

The model uses prior distributions derived from manual data for normal and metastatic components, and a widely spread prior distribution for the non-nodal component, as shown in the top panel of Figure 6. Guided by the prior information, the centres of the three components develop along the directions of the external and internal variables, being updated by the observed data from the individual node. In stage 1, shown in the middle panel, the three groups seem to have fairly well separated means. The normal priors are close to the prior for this group, but the metastatic ones are at a larger distance. The non-nodal group shows a relatively concentrated dispersion compared with its prior distribution. The image configuration at this stage only shows a rough match to the photograph, with some non-nodal spots on the upper left corner being misclassified as metastatic or normal.

In stage 2, with spatial prior taken into account, the means of three groups do not move very much, but isolated pixels are tidied up, and those misclassified non-nodal pixels all recover (see bottom panel of Figure 6). The fitted posterior distributions of the three components in this stage are less concentrated and overlap more with each other, but the resulting image becomes much smoother and its overall structure shows a better match to the photograph of the node.

This example shows that the two-stage algorithm works well in a flexible way. Stage 1 focuses on the distribution convergence with the result of tight fitted groups, and a rough convergence in this stage is enough to generate plausible starting points for the fitting in stage 2. With the MRF spatial prior incorporated in stage 2, although the distribution might not be as concentrated as in stage 1, the image becomes much smoother. We submit that when small groups survive this stage (especially for the case when the metastatic group is small), they are probably real.

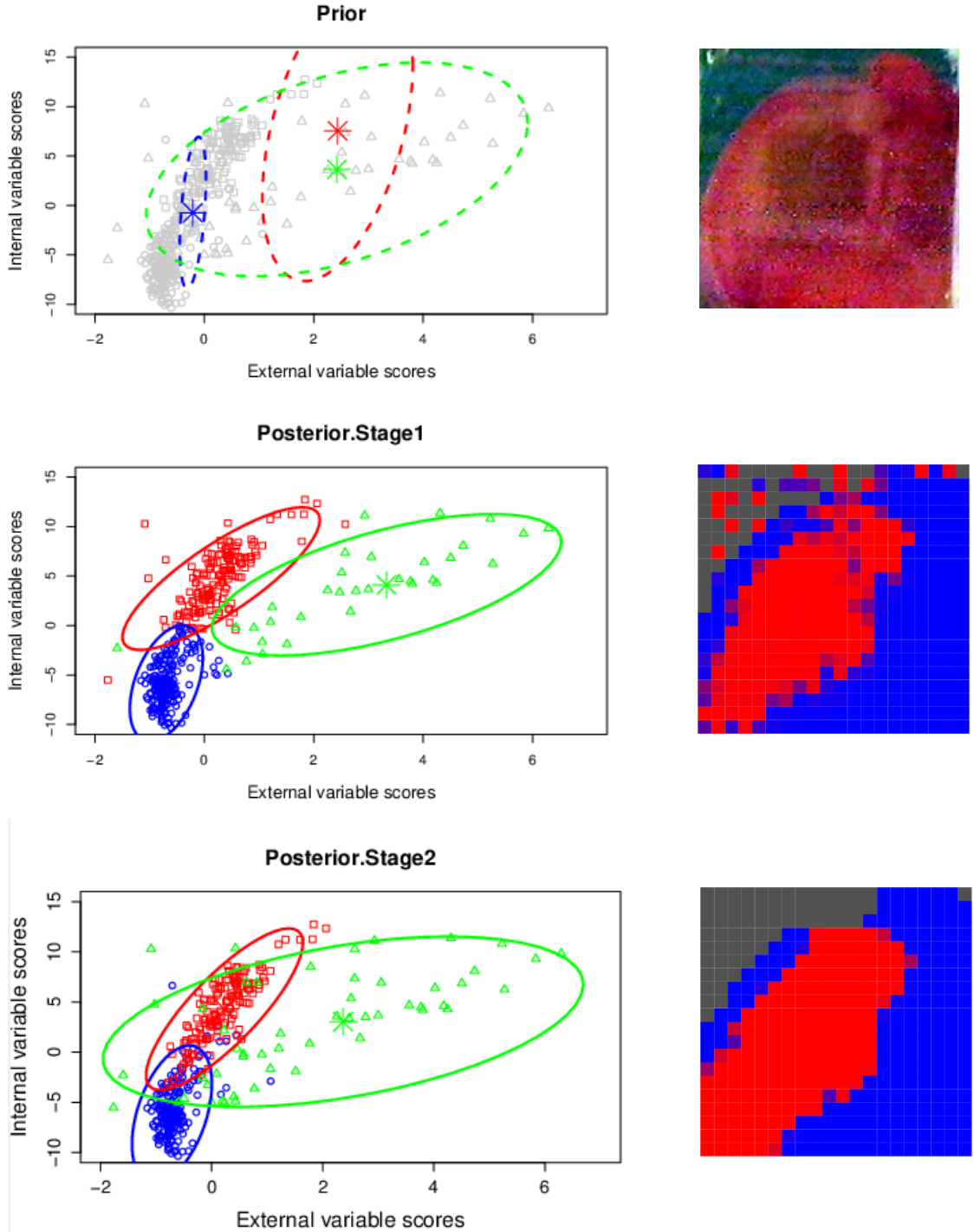

Figure 6: Plot of the two-stage imaging result from a partially metastatic node after discriminant dimension reduction. Two-dimensional prior (top left panel) and posterior (middle and bottom left panels) probability density contour plots showing the effect of stage 1 (middle panel) and stage 2 (bottom panel) model fitting for a mixture of three components (normal in blue, metastatic in red and non-nodal in green). The points show the fitted class membership of each pixel at its current stage, the stars show the prior mean and posterior mean, and the ellipses represent (95%) probability contours of the estimated probability distribution for each component. The right panel shows the photograph of the node and the fitted images from stage 1 and stage 2 with red indicative of metastases, blue of normal, and black of non-nodal spectra.

### 4.3.3 How the two-dimensional prior weight works for image classification model

The prior weights, $K_{jp}$, control the weighting given to the prior distributions in the estimation of the mean and variance of the within-group distributions. The values used in the analysis (see Section 4.1.2) were $K_{1p}$ = diag [5, 2], $K_{2p}$ = diag [3, 1.25], $K_{3p}$ = diag [3.85, 10]. These were arrived at simply by experimentation on the training data, but their relative sizes can be explained. The distributions in Figure 3, which are derived from data on a number of nodes not involved in this analysis, suggest that while the spectra of normal pixels are fairly reproducible from node to node, the metastatic pixels may look different in different nodes: that is, the red distribution looks suspiciously like a mixture. Thus, the fact that the selected weights are stronger for the normal group and weaker for the metastatic makes sense. It also makes sense that for nodal groups the weights for the second component, derived from the node under investigation, should be weaker than those for the first, derived from the same data that were used to inform the prior. For non-nodal group, though from visual inspection of the samples it is very variable from node to node and thus a diffuse prior is used, the fact of using a position parameter $\alpha_{ij}$ increases certainty of the second component derived from the individual node. Hence it is reasonable that the weight for the second component, particularly likely to capture features from non-nodal group, should be stronger than that for the first, derived along the direction for discrimination between the two nodal groups.

### 4.3.4 How the position parameter works for the image classification model

Using a relatively diffuse prior for the non-nodal component may cause a mislabelling problem in the image. However, we have the extra information that the pixels near the edges are more likely to be background, and a position parameter $\alpha_{ij}$ has been incorporated in both stages of the algorithm to exploit this knowledge.

Figure 7 shows an example of a totally normal node with non-nodal pixels mislabelled as metastatic component. This is mainly because the non-nodal component has a similar prior mean for the external variable as the metastatic component, but has a large prior variance. Though prior weight for background is stronger than that for metastatic, without position parameter the non-nodal component is misclassified as metastatic component. The false metastatic pixels are thus incorrectly fitted as shown in the top panel of Figure 7. The second stage algorithm incorporating the spatial correlations develops the smoothness of the mislabelled patches removing the isolated pixels but maintaining the error at the top.

The position parameter introduced in Section 3.2.1.3 can fix the mislabelling problem in the first stage algorithm. By giving higher scores to the pixels on the corner or edge and strong prior weight for non-nodal component it encourages pixels there to be classified as non-nodal. This gives a much better starting configuration for the smooth fitting in stage 2 as shown in the bottom panel of Figure 7.

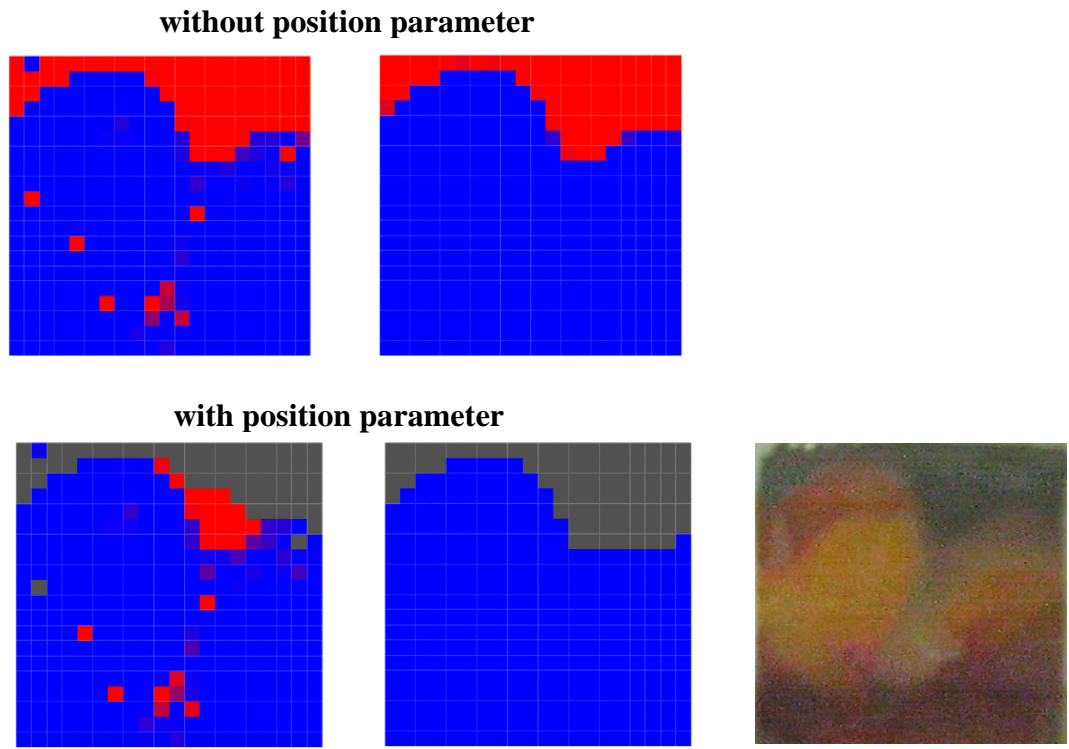

Figure 7: Mapping image of the multivariate mixture model on the spectra from one totally normal node showing the effect of position parameter. The left and middle columns show the images from the 1$^{st}$ and 2$^{nd}$ stage with colour coding at each pixel (red for metastatic, blue for normal and black for non-nodal). The 1$^{st}$ and 2$^{nd}$ rows show the result without and with position parameter incorporated into the model. The right panel of the 2$^{nd}$ row shows a photograph of this node.

## 5. Conclusion and discussions

In this paper a partially supervised image classification algorithm, based on a composite Bayesian multivariate finite mixture model with MRF spatial prior, was developed to represent a scanned node image and to classify scanned nodes as metastatic or normal.

A traditional supervised classification method applied directly to scanned data is not suitable here to derive an algorithm to classify pixels, since there is no reference pathology available for individual pixels in the image and, furthermore, the required training set for the non-nodal group is not available, and these non-nodal areas are highly variable from node to node.

The key issues addressed in this paper are the representation of knowledge and inference methods for using the available knowledge to infer the correct image. The main idea is to enable an integration of *a priori* knowledge from manual data, with accumulated evidence from scanned data, encoded in terms of a joint posterior probability distribution with Markov random field, through a Bayesian formalism.

Before constructing an image classification model, the spectral data are reduced to a two-dimensional space, where the two axes (of external and internal variables) are function-specific and interpretable. Typically the first axis separates the normal and metastatic groups; the second axis allows the model to capture the remaining individual nodal features, particularly from the non-nodal component.

Based on the low-dimensional data, the image classification model is fitted in two stages. In the first stage, a Bayesian multivariate finite mixture model is employed to model three unknown groups (normal, metastatic and non-nodal) in the images guided by an extended normal inverse Wishart prior derived from the manual data. Since the class memberships in the mixture here are not interchangeable, the prior knowledge given here works as an identifiability constraint for normal and metastatic groups. In the second stage, a spatial prior based on a Markov random field (MRF) is then incorporated into the model to represent the continuity of the image. Weak convergence achieved for the EM algorithm in the first stage initializes the RM algorithm in the second stage and leaves more flexibility for model fitting in the second stage. This two-stage approach can, therefore, avoid the common problem that the image classification result by use of EM algorithm is sensitive to initialization due to its property of seeking local maximization. Another advantage of using the composite two-stage approach is that different fitting algorithms are allowed in the two stages, and alternatives could be explored.

A diagonal matrix of prior weights, developed for the extended normal inverse Wishart prior distribution, allows different prior weights in different dimensions. This gives the opportunity to best use information in specific dimensions of the model and thus makes the model more flexible to variability in different dimensions and in these components between nodes. Introducing the two-dimensional prior weight, we lose the simple form for the posterior estimation of $\mu_j$ and $\Sigma_j$ of multivariate $t$ distribution. However, an extra simple iteration introduced in M-step (of both stages) yields fast convergence to update $\mu_j$ and $\Sigma_j$ iteratively.

Being built into the model, the position parameter in both stages works well in recognizing the background component in the mixture model. This helps out some misclassification caused by class-label switching (misclassification) in the first stage algorithm, and can thus provide informative initialization (with correct class labels) for the second stage to enhance the classification accuracy. Different expression forms for constructing the position parameter could be tried to better represent the true feature of the image.

Our results so far suggest that this model can be successful in recognising the variable non-nodal areas automatically and distinguishing between normal and metastatic nodes with the sort of accuracy required to enable clinicians to make a rapid intraoperative diagnosis of sentinel node metastases in breast cancer. The sensitivity of 85% is good enough to reduce the subsequent routine histological examination for those missed metastatic cases during the ESS analysis. The specificity of 94% sounds good suggesting only 6% false positive rate. However, there may still some scope for improving on it when examining the population positive predictive value (PPV). PPV measures the probability of a patient with a positive test actually having the disease of interest and is often of more interest for the clinicians in the long-term assessment of the model. Assuming a population prevalence of 20%, the same values of sensitivities and specificities would lead to a PPV of 78%. This means that 22% of the cases detected as positive are in fact false positives, leading to unnecessary surgical axillary dissection. In this study, considering small training data set, it would be difficult to obtain a high PPV above 90% and thus maximizing specificity is still a reasonable choice by giving a more stable diagnostic result. In the future, it is of our interest to improve the PPV by carrying out a prospective clinical study and analysis on a larger number of patients.

As a general method, this image classification model may also be applied to many other situations for both noise/background recognition and multi-group tissue classification, when the group feature information is only available for the groups of interest.

**Acknowledgements**

The authors acknowledge research funding from the U.S. Department of Defense Breast Imaging Program (Award No. W81XWH-04-1-0589) and from Hamamatsu Photonics, Hamamatsu, Japan. The authors are grateful to the Peacock Trust, the National Institutes of Health (NIH) Network for Translational Research into Optical Imaging (NTROI) program (U54 CA104677), Experimental Cancer Medicine Centre (ECMC), Comprehensive Biomedical Research Centre (CBRC), and the Academic Research Fund (AcRF: RI 6/14 ZY) of National Institute of Education, Nanyang Technological University, Singapore, for their support of this work at University College London (UCL), UCL hospital and NTU.